\documentclass{aa} % for a referee version
\usepackage{caption}
\usepackage{subcaption}
\usepackage{graphicx}
\usepackage{float}
\usepackage{txfonts}
\begin{document}

   \title{Pseudostreamer influence on flux rope evolution}

   \author{A. Sahade
          \inst{1,2,3}
          \and
          M. C\'ecere\inst{1,3}
          \and
          M.V. Sieyra\inst{4}
          \and
          G. Krause\inst{5,6}
          \and
         H. Cremades\inst{7}
          \and
         A. Costa\inst{1}
          }

   \institute{Instituto de Astronom\'{\i}a Te\'orica y Experimental, CONICET-UNC, C\'ordoba, Argentina.
         \and
            Facultad de Matem\'atica, Astronom\'{\i}a, F\'{\i}sica y Computaci\'on, Universidad Nacional de C\'ordoba (UNC), C\'ordoba, Argentina.
         \and
             Observatorio Astron\'omico de C\'ordoba, UNC, C\'ordoba, Argentina. 
        \and
             Centre for mathematical Plasma Astrophysics, Department of Mathematics, KU Leuven, Celestijnenlaan 200B, B-3001 Leuven, Belgium 
        \and
             Instituto de Estudios Avanzados en Ingenier\'{\i}a y Tecnolog\'{\i}a, CONICET, C\'ordoba, Argentina.
        \and
             Facultad de Ciencias Exactas, F\'{\i}sicas y Naturales, UNC, C\'ordoba, Argentina.
        \and
             Grupo de Estudios en Heliof\'isica de Mendoza, CONICET, Universidad de Mendoza, Mendoza, Argentina. 
             }

   \date{Received ; accepted }

% \abstract{}{}{}{}{} 
% 5 {} token are mandatory
 
  \abstract
  % context heading (optional)
  % {} leave it empty if necessary  
   {A highly important aspect of solar activity is the coupling between eruptions and the surrounding coronal magnetic field topology, which determines the trajectory and morphology of the event.
   Pseudostreamers (PSs) are coronal magnetic structures formed by arcs of twin loops capped by magnetic field lines from coronal holes of the same polarity that meet at a central spine. They contain a single magnetic null point in the spine, just above the closed field lines, which potentially influences the evolution of nearby flux ropes (FRs).}
  % aims heading (mandatory)
   {Because of the impact of magnetic FR eruptions on space weather, we aim to improve current understanding on the deflection of CMEs. To understand the net effect of the PS on FR eruptions is first necessary to study diverse and isolated FR-PS scenarios, which are not influenced by other magnetic structures.
}
  % methods heading (mandatory)
   {We performed numerical simulations in which a FR structure is in the vicinity of a PS magnetic configuration. The combined magnetic field of the PS and the FR results in the formation of two magnetic null points. We evolve this scenario by numerically solving the magnetohydrodynamic equations in 2.5D. The simulations consider a fully ionised compressible ideal plasma in the presence of a gravitational field and a stratified atmosphere. 
   }
  % results heading (mandatory)
   {We find that the dynamic behaviour of the FR can be categorised into three different classes based on the FR trajectories and whether it is eruptive or confined. Our analysis indicates that the magnetic null points are decisive in the direction and intensity of the FR deflection and their hierarchy depends on the topological arrangement of the scenario. Moreover, the PS lobe acts as a magnetic cage enclosing the FR. We report that the total unsigned magnetic flux of the cage is a key parameter defining whether the FR is ejected or not.  
   }
  % conclusions heading (optional), leave it empty if necessary 
   {}


   \maketitle

% --------------------------------------------------------------------   
% --------------------------------------------------------------------     
\section{Introduction}
% --------------------------------------------------------------------   
% --------------------------------------------------------------------        

Magnetic flux ropes (FRs) are thought to be the central structures of solar eruptions, including prominence/filament eruptions, flares, and coronal mass ejections (CMEs). The physical process connecting these phenomena is the eruption of the magnetic flux rope system \citep{Zhang2001,vanDriel2015,Green2018,Jiang2018,Yang2018,Filippov2019}. Knowing whether the FR will erupt or not is, naturally, fundamental to predicting a CME event. \citet{Gronk2016IAUS..320.} pointed out several mechanisms that can decelerate and confine eruptions in the corona. The first one is the action of gravity, which prevents the eruption when the energy of the FR is not enough to escape the gravitational potential of the Sun \citep{Filippov2021}. Even if the FR has the energy to escape gravity, the eruption would be confined if the overlying arcade field, whose lines form a magnetic cage, is too strong or do not quickly decay with height
\citep{Torok2005,WangyZhang2007,Chen2013ApJ...778...70C,Amari2018,Baumgartner2018,Jing2018}. Recently, \citet{Li2020ApJ...900..128L,Li2021ApJ...917L..29L,Li2022ApJ...926L..14L} proved that there exists a negative correlation between the flare eruptivity (i.e. if it has an associated CME) and the total unsigned magnetic flux of the active region producing the flare, which describes the strength of the background field confinement. 

Once the eruption occurs, it is essential to understand the path that a CME will follow in order to predict its geoeffectiveness. This requires knowledge of any non-radial propagation of the CME, for which deflections in the trajectory must be studied. It is widely known that the magnetic structures in the vicinity of FRs are capable of deflecting them both in latitude and longitude. While coronal holes \citep[e.g.,][]{2006AdSpR..38..461C,2009JGRA..114.0A22G,Sahade2020,Sahade2021} and active regions \citep[e.g.,][]{2015ApJ...805..168K,Mostl2015,Wang2015} deflect FRs against their location, heliospheric current sheets \citep[e.g.,][]{2015SoPh..290.3343L,Wang2020JGRA}, helmet-streamers \citep[e.g.,][]{Zuccarello2012,Yang2018} and pseudostreamers (PSs) \citep[e.g.,][]{Bi2013,Wang2015PS,cecere2020,Wang2020JGRA} attract FRs to their low magnetic energy regions. Combined effects of the several structures at different heights can be seen in, for example, \citet{Sieyra2020}. In previous studies \citep{Sahade2020,Sahade2021}, we found that the presence of a coronal hole nearby the eruptive region forms a magnetic null point that attracts the FR. The null point can be located between the FR and CH (case of anti-aligned polarities) or at the other side of the FR (case of aligned polarities). The first scenario produces an initial deflection towards the CH and a second deflection against its position. On the other hand, the aligned polarities cases lead to a single deflection in which the FR moves away from the CH. 
All the final paths are opposite to the location of the coronal hole by the ``channelling'' of the magnetic field lines, i.e., the FR is guided to follow the least resistance path. \citet{Mostl2015} and \citet{Wang2015} studied an event on 2014 January 7 whose deflection seems to be caused by the magnetic pressure gradient from a nearby active region and whose final path is also channelled by the configuration of the magnetic field lines to the least resistance direction. \citet{shen2011} concluded that the trajectory in the early stages is influenced by the background magnetic energy gradients, inducing the CME to propagate towards the region with the lowest magnetic energy density. Similar results were found by \citet{Sieyra2020} where most of the analysed CMEs eruptions were aligned with the direction of the magnetic energy decrease. It also showed that most of the deflection occurs at heights lower than 2.4\,$R_{\odot}$, suggesting that it is of utmost importance to study the trajectory in the early stage.

Streamers are characterised by containing a region of null magnetic energy, therefore they can act as a potential well attracting CMEs towards them \citep{Kay2013}. In particular, the PS contains a single magnetic null point above the closed field lines. These closed field lines that overlie two (or an even number of) polarity inversion lines, are covered by open field lines of the same polarity, without a current sheet, forming the spine of the pseudostreamer  \citep{2014ApJ...787L...3R}. 
Observational studies have suggested that there is a null point hierarchy: the rolling motions and deflections of prominences are caused by the nearest local null point and the CMEs move in a non-radial direction towards the global null point located at higher altitudes associated to helmet streamers or pseudostreamers \citep{2013SoPh..287..391P}. In order to explain the physical processes involved in the deflection of the eruptive phenomenon there are some numerical studies that analyse the CME deflections in presence of PS structures. For example, \citet{Zuccarello2012} found, in a scenario where the heliospheric current sheet and PS are both present, a CME that erupts from one of the PS lobes and is firstly deflected towards the null point of the PS and then continues moving towards the heliospheric current sheet. A similar behaviour is found in the simulation performed by \citet{wyper2021}, in which the PS is embedded in a helmet streamer. Recently, \citet{Karna2021} modelled the eruptive filament observed on 2015 April 19, which was embedded in a lobe of the pseudostreamer and was directed towards the PS null point. 
Although it is well established, both numerically and observationally, that FRs inside PS deflect towards the PS spine \citep[e.g.,][]{Torok2011,Zuccarello2012,Yang2015ApJ,Karna2021}, to date there are no studies that analyse how the trajectory of a FR is affected by variations in the FR-PS configuration. 

In this paper, we model cases where only one FR interacts with the PS structure, and we analyse its influence on the FR trajectory at low coronal heights by 2.5 MHD numerical simulations. In Section 2, we describe the numerical model details and parameters for the presented cases. In Section 3, we present results arising from the several simulations performed. We observed that the FR-PS interactions can be distinguished into three separate classes that exhibit differences in the dynamic behaviour of the FR. On one hand, differences in the magnetic field topology lead to different hierarchies of the null points and consequently changes in the dynamical behaviour. On the other hand, we found that the eruption or confinement of the FR strongly depends on the unsigned magnetic flux of the magnetic cage. Discussion and final remarks are presented in Section 4.
 
% --------------------------------------------------------------------   
% --------------------------------------------------------------------     
\section{Numerical simulations}
% --------------------------------------------------------------------   
% --------------------------------------------------------------------     
To study the interaction between a FR and a PS, we present a scenario where both structures interact in isolation. In this way, we avoid the possible effects of other magnetic structures that could affect the FR behaviour, allowing a comprehensive analysis of the PS influence in the FR evolution. We consider the ideal MHD equations in presence of a gravitational field to solve the 2.5 dimensional model.
In CGS units in the Cartesian conservative form we have:

\begin{align}
    &\frac{\partial\rho}{\partial t}+\nabla\cdot(\rho\vec{v})=0 &(continuity)&  \label{e:cont} \\
    &\frac{\partial (\rho \vec{v})}{\partial t} + \nabla \cdot \left(\rho \vec{v} \vec{v} - \frac{1}{4\pi} \vec{B}\vec{B} \right) + \nabla p + \nabla\left( \frac{B^2}{8\pi}\right)   = \rho \vec{g}  &(momentum) \label{e:euler} \\
    &\frac{\partial E}{\partial t} + \nabla \cdot \left[\left(E + p + \frac{B^2}{8\pi}\right)\vec{v} -\frac{1}{4\pi} \left(\vec{v\cdot B}\right)\vec{B}\right] = \rho \vec{g v}  &(energy) \label{e:consE} \\
    &\frac{\partial \vec{B}}{\partial t} + \vec{\nabla \cdot} \left(\vec{v} \vec{B} - \vec{B} \vec{v} \right) = \vec{0} &(induction) \label{e:induccion}
\end{align}

\noindent
where $\rho$ represents the plasma density, $p$ the thermal pressure, $\vec{v}$ the velocity, $\vec{B}$ the magnetic field, and $\vec{g}$ the gravity acceleration. $E$ is the total energy (per unit volume), given by  
\begin{equation*}
    E = \rho \epsilon + \frac{1}{2} \rho v^2 + \frac{B^2}{8\pi},
\end{equation*}
where $\epsilon$ is the internal energy and
\begin{equation*}
{\vec{j}=\frac{c}{4\pi}{\nabla\times}\vec{B}, } \,
\end{equation*}
is the current density, with $c$ being the speed of light.

In addition to the MHD equations, the divergence-free condition of the magnetic field must be fulfilled, i.e.
\begin{equation}\label{e:divB}
 \vec{\nabla\cdot} \vec{B} = 0\, .
\end{equation}

We assume that the medium is a fully ionised hydrogen plasma, for which is valid the perfect gas law $p = 2\rho k_B T/m_i = (\gamma - 1) \rho \epsilon$. $k_B$ is the Boltzmann constant, $T$ the plasma temperature, $m_i$ the proton mass, and $\gamma = 5/3$ the specific heat relation. 

Simulations were performed using the FLASH Code \citep{2000ApJS..131..273F} in its fourth version, operated under an adaptive refinement mesh with the USM (Unsplit Staggered Mesh) solver, which uses a second-order directionally unsplit scheme with a MUSCL-type (Monotonic Upstream-centered Scheme for Conservation Laws) reconstruction. We use the local Lax-Friedrichs Riemann solver, which is a diffusive solver providing the necessary dissipation to emulate the magnetic resistivity and use the ideal MHD equations \citep{Sahade2020}.
Outflow conditions (zero-gradient) are used at lateral and upper boundaries, while the line-tied condition is used at the lower boundary, which imposes the condition of null velocity and constant magnetic field for the ghost cells \citep{1987SoPh..114..311R}. In the guard-cells of the boundary the magnetic field is linearly extrapolated to preserve the divergence-free configuration.  
The highest resolution corresponds to $\sim[0.1 \times 0.1]~\mathrm{Mm}^2$ cells, in a $[-700,700]~\mathrm{Mm}\times [0,700]~\mathrm{Mm}$ physical domain, where pressure and temperature gradients satisfy the refinement criterion.

% --------------------------------------------------------------------   
% --------------------------------------------------------------------     
\subsection{FR and PS magnetic model}\label{ss:magneticmodel}

The  modelling of the FR magnetic structure is based on the catastrophe model by \citet{1990JGR....9511919F}  consisting of an out-of equilibrium magnetic configuration that  triggers the FR ejection. The model of the  PS is based on the magnetic configuration proposed by \citet{Edmondson10}. However, to better reproduce the decay of the magnetic field with altitude in the solar corona, we replace the constant background magnetic field in the $y$-direction by an exponentially decaying field. The 
$x$-direction is oriented along the horizontal coordinate, the $y$-direction corresponds to the vertical coordinate and the $z$-direction is the direction of symmetry. Combining both models, the total magnetic field is given by:
\begin{align*}
    B_x&=B_{x,\mathrm{FR}}+B_{x,\mathrm{PS}} \, ,\nonumber \\
    B_y&=B_{y,\mathrm{FR}}+B_{y,\mathrm{PS}} \, , \nonumber \\
    B_z&=B_{z,\mathrm{FR}}\, .
\end{align*}
The magnetic field components of the FR are given by the sum of a current wire, an image current wire and a line dipole:
\begin{align}\label{e:BFR}
    B_{x,\mathrm{FR}}=-&B_\phi(R_-)\tfrac{(y-h_0)}{R_-} +  B_\phi(R_+)\tfrac{(y+h_0)}{R_+} - \nonumber \\ &MdB_\phi{\scriptstyle\left(r+\tfrac{\Delta}{2}\right)}\left(r+\tfrac{\Delta}{2}\right)\tfrac{x^2-(y+d)^2}{R_d^4} \, ,\nonumber \\
    B_{y,\mathrm{FR}}=\quad &B_\phi(R_-)\tfrac{x}{R_-} - B_\phi(R_+)\tfrac{x}{R_+}-\nonumber \\
    & MdB_\phi{\scriptstyle\left(r+\tfrac{\Delta}{2}\right)}\left(r+\tfrac{\Delta}{2}\right)\tfrac{2x(y+d)}{R_d^4}  \, , \nonumber \\
    B_{z,\mathrm{FR}}=\quad &B_{\text{z}}(R_-)\, .
\end{align}

\noindent
In these expressions, $h_0$ is the initial height of the FR, $M$ is the intensity of the line dipole at depth $d$, $r$ is the current wire radius, $\Delta$ is the thickness of the transition layer between the current wire and the exterior, and $R_\pm = \sqrt{x^2+(y\pm h_0)^2}$ and $R_d = \sqrt{x^2+(y+d)^2}$ are the distances taken from different origins (image and current wire, and dipole, respectively). Also, 
\begin{equation} \label{e:Bphi}
 B_\phi{(R)}\!=\!
    \left\{
\begin{array}{rl}
\begin{alignedat}{2}
&\tfrac{2\pi}{c}j_0R  && 0\leq R < r-\frac{\Delta}{2}\\
&\tfrac{2\pi j_0}{cR}\left\{\tfrac{1}{2}\left(r-\tfrac{\Delta}{2}\right)^2-\left(\tfrac{\Delta}{2}\right)^2 +\right. 
\\
&\tfrac{R^2}{2}+\tfrac{\Delta R}{\pi}\text{sin}\left[\tfrac{\pi}{\Delta}\left(R-r+\tfrac{\Delta}{2}\right)\right]+ \qquad && r-\frac{\Delta}{2}\!\leq \!R\! <r+\frac{\Delta}{2}
\\
&\left.\!\!\!\left(\tfrac{\Delta }{\pi}\right)^2\cos\left[\tfrac{\pi}{\Delta}\left(R-r+\tfrac{\Delta}{2}\right)\right]\right\} \\
&\tfrac{2\pi j_0}{cR}\left[r^2+\left(\tfrac{\Delta}{2}\right)^2-2\left(\tfrac{\Delta}{\pi}\right)^2\right]  && r+\frac{\Delta}{2} \leq R
\end{alignedat}
\end{array} \right.
\end{equation}
\begin{equation} \label{e:jz}
  j_z{(R)}\!=\!
    \left\{
\begin{array}{rl}
\begin{alignedat}{2}
&\!j_0 &&0\leq R < r-\frac{\Delta}{2}\\
&\!\tfrac{j_0}{2}\left\{\cos\left[\tfrac{\pi}{\Delta}\left(R-r+\tfrac{\Delta}{2}\right)\right]+1\right\} \quad  && r-\frac{\Delta}{2}\!\leq \!R\! <r+\frac{\Delta}{2}\\
&\! 0  && r+\frac{\Delta}{2} \leq R
\end{alignedat}
\end{array} \right.
\end{equation}
\noindent
where $j_0$ is a current density. The component $B_\mathrm{z}$ of the magnetic field and the current distribution $j_{\phi}$, are described by:
\begin{equation}\label{e:Bfieldz}
    B_\mathrm{z}(R) = \tfrac{4\pi j_1}{c}\sqrt{\left(r-\tfrac{\Delta}{2}\right)^2-R^2}\, , 
\end{equation}
\begin{equation}\label{e:jphi}
    j_\phi(R) = j_1R\left[\sqrt{\left(r-\tfrac{\Delta}{2}\right)^2-R^2}\right]^{-1}\,, 
\end{equation}
where $j_1$ is a current density. These expressions are valid in $0\leq R < r-\frac{\Delta}{2}$ and are null in the rest of the domain. 

The magnetic field components of the PS are composed by a line dipole and a potential field:
\begin{subequations}
\label{eq:optim}
\begin{align}\label{e:BfieldPSx}
    B_{x,\mathrm{PS}}(x,y)=&\frac{2 \sigma  B_\text{PS} (x - x_\text{PS}) (y - y_\text{PS})}{((x - x_\text{PS})^2 + (y - y_\text{PS})^2)^2} + \\ \nonumber
    & B_0\ \sin\left(\frac{x-x_\text{PS}}{H}\right)\, \exp[-y/H]\, ,   \\ \nonumber
\end{align}
\begin{align} \label{e:BfieldPSy}
    B_{y,\mathrm{PS}}(x,y)= & - \frac{2\sigma B_\text{PS} (x - x_\text{PS})^2}{((x - x_\text{PS})^2 + (y - y_\text{PS})^2)^2} + \\ \nonumber
    & \frac{\sigma B_\text{PS}}{(x - x_\text{PS})^2 + (y - y_\text{PS})^2} +  \\ \nonumber
    & B_0\ \cos\left(\frac{x-x_\text{PS}}{H}\right)\,\exp[-y/H] \, , 
\end{align}
\end{subequations}
where $\sigma B_\mathrm{PS}$ is the strength of the magnetic field due to a single line dipole ($\sigma =2\times10^{21}$ is a dimensionless scale factor) positioned at $(x, y) = (x_\mathrm{PS}, y_\mathrm{PS})$, $B_0$ is the strength of the background field at $(x, y) = (x_\mathrm{PS}, 0)$, and $H=600$ Mm is the height decaying factor.

\begin{figure}
\centering
    \includegraphics[width=0.9\linewidth]{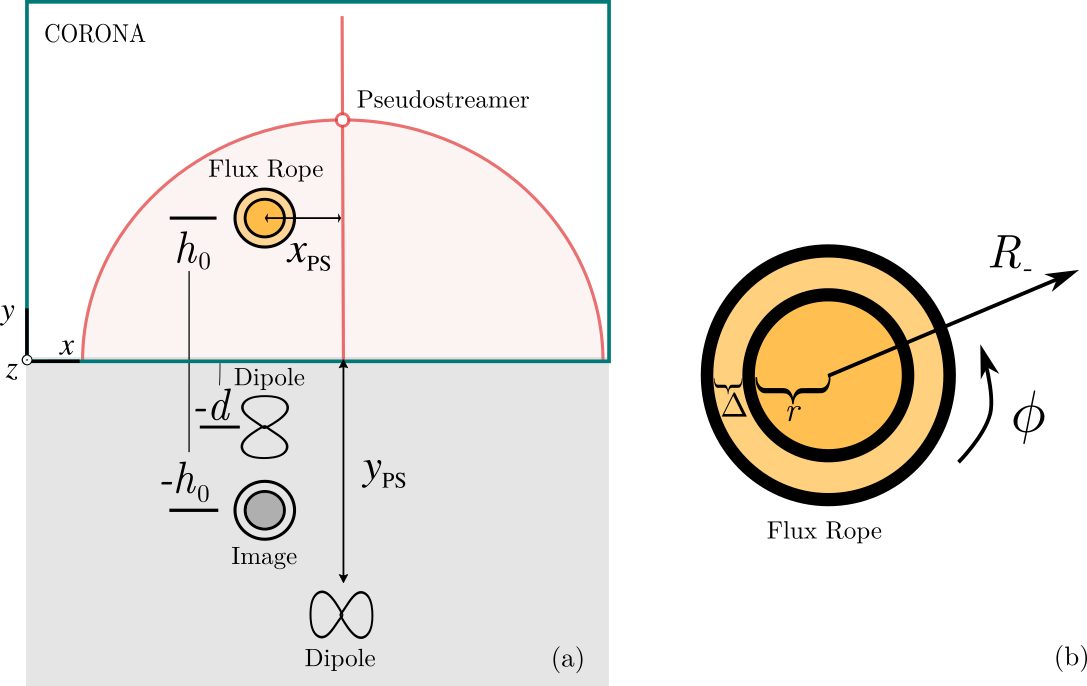}
     \caption{(a) Flux rope (orange circle) and pseudostreamer (red structure) scheme, not to scale. The green frame indicates the simulated region. $h_0$ is the FR height and $x_\mathrm{PS}$ is the distance from the FR to the PS spine (red vertical line), whose height is parametrised by $y_\mathrm{PS}$. The line dipole and the image current are located at depth $d$ and $h_0$, respectively. (b) Internal structure of the FR: $r$ is the radius, $\Delta$ is the thickness of the transition layer, $R_{-}$ is the radial coordinate from the FR centre, and $\phi$ the poloidal coordinate.}
    \label{f:scheme}
\end{figure}

Figure \ref{f:scheme}(a) shows a scheme with the distribution of the magnetic structures and \ref{f:scheme}(b) the internal structure of the FR. In the left panel, the green frame (top) represents the simulated domain and the grey shaded area (bottom) contains the magnetic components that are out of the simulation box. The PS model produces a four-flux system, the separation between these regions (red lines) is characterised by a magnetic null point (red circle). Two of the fluxes have a closed topology (red shaded area) and form the PS lobes that are divided by the spine (vertical red line). Outside this region , delimited by the semicircular red line, two open fluxes of equal polarity converge towards the spine and surround the PS structure.

% --------------------------------------------------------------------   
% --------------------------------------------------------------------     
\subsection{Thermodynamic variables}

We simulate the solar atmosphere by adopting a multi-layer structure \citep{Mei2012}. The chromosphere lies between the heights $0 \leq y \leq h_\mathrm{ch}=10\,$Mm with constant temperature $T_\mathrm{ch}=10\,000\,\text{K}$. The base of the corona is at a height $h_\mathrm{c}=15\,\text{Mm}$ and has constant temperature $T_\mathrm{c}=10^6\,\text{K}$. These layers are connected by the transition region, which has a linearly increasing temperature. Thereby, the initial temperature distribution is given by 
\begin{equation}
    {\textstyle T(y)=}
    \left\{
\begin{array}{rl}
\begin{alignedat}{2}
&{\textstyle T_\mathrm{ch}} && 0\leq y < h_\mathrm{ch}\\
&{\textstyle (T_\mathrm{c}-T_\mathrm{ch})\left[\frac{y-h_\mathrm{ch}}{h_\mathrm{c}-h_\mathrm{ch}}\right]+T_\mathrm{ch}} \quad && h_\mathrm{ch}\leq y < h_\mathrm{c}\\
& {\textstyle T_\mathrm{c}}  && h_\mathrm{c} \leq y.
\end{alignedat}
\end{array} \right.  
\end{equation}

The temperature inside the FR ($T_{\text{\tiny{FR}}}$) varies according to the following temperature distribution:
\begin{equation}
    \!T(R_-)\!=\!
    \left\{
\begin{array}{rl}
\begin{alignedat}{2}
&\!T_{\text{\tiny{FR}}} && 0\leq R < r-\frac{\Delta}{2}\\
&\!\!(T_\mathrm{c}\!-\!T_{\text{\tiny{FR}}})\!\left[\tfrac{R_--(r+\Delta/2)}{\Delta}\right]\!+\!T_{\text{\tiny{FR}}} \quad && r-\frac{\Delta}{2}\!\leq \!R\! <r+\frac{\Delta}{2}\\
&\! T_\mathrm{c} && r+\frac{\Delta}{2} \leq R.
\end{alignedat}
\end{array} \right.
\end{equation}

We consider a current-free atmosphere in hydrostatic equilibrium. Hence, the background pressure $p(y)$ is only a function of height considering a system having the $y$-axis aligned to the gravity acceleration  (i.e., $\vec{g} =\frac{-G M_\sun}{(y + R_\sun)^2}\vec{e}_y$, where $G$ is the gravitational constant, $M_\sun$ is the solar mass, $R_\sun$ is the solar radius, and $y = 0$ corresponds to the solar surface). Therefore, the atmospheric pressure is:
 
\begin{equation}\label{pres}
    {\textstyle  p(y)=\!}
    \left\{
\begin{array}{rl}
\begin{alignedat}{2}
&{\textstyle \!\!p_\mathrm{ch}\exp{\! \left[\frac{\alpha}{T_\mathrm{ch}}\left(\frac{1}{h_\mathrm{ch}+R_{\sun}}-\frac{1}{y+R_{\sun}}\right)\right]}} && 0\leq y < h_\mathrm{ch} \\
&{\textstyle\! \!p_\mathrm{ch}\exp{\!\left[-\int_{h_\mathrm{ch}}^{y}\frac{\alpha}{T{\scriptstyle(y')}}(R_{\sun}+y')^{-2} dy'\right]}} \quad &&  h_\mathrm{ch}\leq y < h_\mathrm{c}\\
&{\textstyle \! \!\frac{k_B}{N_Am_i}T_\mathrm{c}n_\mathrm{c}\exp{\!\left[-\frac{\alpha}{T_\mathrm{c}}\left(\frac{1}{h_\mathrm{c}+R_{\sun}}-\frac{1}{y+R_{\sun}}\right)\right]}}  &&  h_\mathrm{c} \leq y ,

\end{alignedat}
\end{array} \right.  
\end{equation}
where 
\begin{equation*}
    {\textstyle p_\mathrm{ch}(y)=\frac{k_B}{N_Am_i}T_\mathrm{c}n_\mathrm{c}\exp{\left[\int_{h_\mathrm{ch}}^{h_\mathrm{c}}\frac{\alpha}{T(y')}(R_{\sun}+y')^{-2} dy'\right]}} \, ,
\end{equation*}
The number density at height $y = h_\mathrm{c}$ in the corona is $n_\mathrm{c}=3\times10^8$, $\alpha= \frac{m_i G M_\sun}{2k_B}$, and $N_A$ is the Avogadro number.
The internal pressure of the FR is obtained by proposing a solution close to the equilibrium:
\begin{align} \label{e:presion}
    p_\text{\tiny{FR}}(x,y) = p(y)& +\tfrac{1}{c}\int_{R}^{r+\frac{\Delta}{2}}B_\phi{\scriptstyle(R')}j_z{\scriptstyle(R')}dR'\nonumber\\
    &-\tfrac{1}{c}\int_{R}^{r+\frac{\Delta}{2}}B_{\text{z}}{\scriptstyle(R')}j_\phi{\scriptstyle(R')}dR'.
\end{align}

\noindent
The associated plasma densities are obtained from the adopted equation of state, i.e.:
\begin{equation}
{\textstyle \rho=\frac{m_i p(y)}{2k_BT(y)}}.    
\end{equation}

% --------------------------------------------------------------------   
% --------------------------------------------------------------------     
\subsection{Setup} \label{setup}
% --------------------------------------------------------------------   
% --------------------------------------------------------------------  

\begin{table}[]
    \centering
        \caption{PSs parameters and class of interaction between the FR and PSs.}
    \begin{tabular}{r| r | r | r | l | l  }
         &$B_0$[G] & $B_\text{PS}$[G] & $x_\text{PS} $[Mm]&$y_\text{PS} $[Mm] & Class   \\
        \hline
        PS1-L & $1$ & $-1.284$ &$210$ &$-360$& I$_\mathrm{e}$ \\   % b1y280x2fD
        PS1-C & $1$ & $-1.284$ &$140$ &$-360$& I$_\mathrm{e}$ \\   % b1y280x2fC
        PS1-R & $1$ & $-1.284$ &$70$ &$-360$& I$_\mathrm{e}$  \\   % b1y280x2fI
        \hline
        PS2-L & $2 $ & $-2.569$ &$210$ &$-360$& I$_\mathrm{ne}$ \\ % b2y280x2fD
        PS2-C & $2 $ & $-2.569$ &$140$ &$-360$& I$_\mathrm{ne}$ \\ % b2y280x2fC
        PS2-R & $2 $ & $-2.569$ &$70$ &$-360$& I$_\mathrm{e}$   \\ % b2y280x2fI
        \hline
        PS3-L & $0.5$ & $-0.203$ &$210$ &$-180$& O \\ % b05y140x2fD
        PS3-C & $0.5$ & $-0.203$ &$140$ &$-180$& O \\ % b05y140x2fC
        PS3-R & $0.5$ & $-0.203$ &$70$ &$-180$& O  \\ % b05y140x2fI

    \end{tabular}
    \tablefoot{Parameter $B_0$ determines the magnetic field strength that surrounds the PS. $B_\text{PS}$ modulates the magnetic field strength inside the PS lobes. The parameters  $x_\text{PS}$ and $y_\text{PS}$ indicate the position of the dipole that produces the PS lobes. The Class column indicates the initial scenario (I=inner, O=outer) and if the FR erupts or not (subscripts `e' and `ne', respectively).}
    \label{t:PSparameters}
\end{table}

\begin{figure*}
    \centering
    \includegraphics[width=\linewidth]{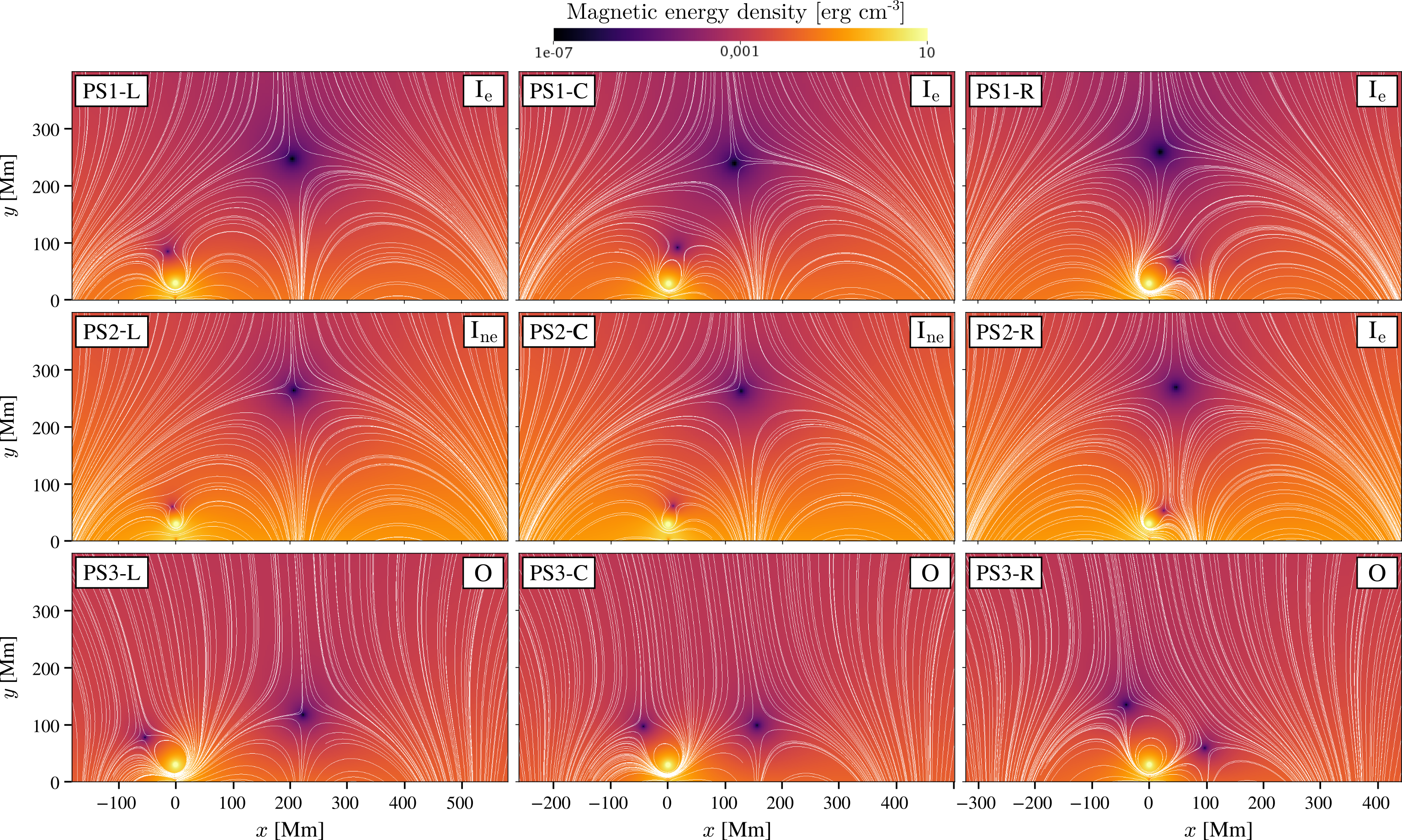}
    \caption{Initial scenarios for simulations listed in Table 1. The case name and the class of interaction are indicated in  boxes to the left and right of each panel, respectively. The colour represents the magnetic energy density, null points can be noticed in dark violet. The magnetic field lines are drawn in white. }
    \label{f:scenarios}
\end{figure*}   

We perform several simulations to analyse the evolution of a FR interacting with different PS configurations. For all cases, we establish a single FR configuration and model the different cases by varying the parameters describing the magnetic structure of the PS. The simulated FR is warm, its temperature equals the coronal one ($T_{\mathrm{FR}}=1~\mathrm{MK}$), it has an initial height of $h_0=30~\mathrm{Mm}$, a radius of $r=2.5~\mathrm{Mm}$ and a transition layer thickness of $\Delta=0.25~\mathrm{Mm}$. Its magnetic parameters are  $j_0=435~\textrm{statA cm}^{-2}, j_1=322~\textrm{statA cm}^{-2}$, $M=1$ and $d=-3.125~\textrm{Mm}$.

We list in Table~\ref{t:PSparameters} the parameters of the selected PSs and, in the last column, the resulting interaction class with the FR. To fix the height of the null point ($y_n$) we determine the parameter $B_{\textrm{PS}}$ by making zero equation \eqref{e:BfieldPSy} ($B_{y,\mathrm{PS}}(x_\mathrm{PS},y_n)=0$). PS1 and PS2 cases correspond to PSs with the null point at a height of $\sim 280$ Mm and lobes of $\sim 400$ Mm wide. PS3 cases have the null point at height $\sim 140$ Mm and their lobes width is $\sim 200$ Mm. Figure~\ref{f:scenarios} shows the magnetic energy density and field lines for each case listed in Table~\ref{t:PSparameters}. The FR is located to the left of the PS spine. The nomenclature L (left), C (centred) and R (right) indicates the alignment of the FR respect to the left PS lobe.These cases are representative of a larger sample of performed simulations. They cover the following combination of parameters: background magnetic field $B_0=\{0.5,1,2\}$ G; null point height $y_n=\{140,280\}$ Mm; lobe width $w\sim\{y_n,1.5 y_n\}$ and FR alignment $\{$R, C, L$\}$. The coupling of the PS and FR magnetic fields alters the PS shape, resulting in a displacement of the PS null point and the appearance of a new null point for all cases. We refer to the PS null point, which is located at a certain height along the spine, as global null point (GNP). Likewise, we name the new null point, produced by the addition of the FR, local null point (LNP).
 
Two initial scenarios are possible when the PS and FR magnetic field are combined. In one scenario, the LNP is closer to the FR and it forms inside the PS structure. The magnetic field lines of the PS lobe that overlay the FR form a confining cage. In the other scenario, the FR field is strong enough to change the PS topology, bending the left lobe field, and the LNP is associated with a new spine-like structure outside the PS. Therefore, there is no arcade over the FR forming a magnetic cage. We refer to the interaction resulting from the first scenario as class I (inner; top and middle rows of Fig.~\ref{f:scenarios}), and the second scenario results in a class O (outer) interaction (bottom row of Fig.~\ref{f:scenarios}). Class I cases can also be divided according to whether they are eruptive or non-eruptive events (see subscripts $e$, for eruptive, and $ne$, for non-eruptive, in Table~\ref{t:PSparameters} and Fig.~\ref{f:scenarios}). As we show below, the O class has only eruptive events, possibly because the rearrangement of the topology favours the ejection. We focus this work in the description and analysis of the FR dynamic behaviour and evolution according to the interaction class.

% --------------------------------------------------------------------   
% --------------------------------------------------------------------    

\section{Results}
% --------------------------------------------------------------------   
% --------------------------------------------------------------------     
We present two different analyses: a dynamical and a quantitative one. In the first one, we analyse and compare how the FR trajectories are influenced by the presence of the LNP and the GNP. For the second one, we study the forces that are involved in the dynamics, the unsigned magnetic flux of the magnetic cage, and how the FRs are affected by these factors.

\subsection{ Dynamic behaviour analysis}

In this section we analyse the similarities and differences in the FR deflection depending on the class (I or O) and on whether the event is eruptive or non-eruptive. To facilitate the interpretation of the trajectories we use a new reference frame defined as $x'=x-x_\mathrm{GNP}$ centred in the GNP of each PS, therefore all the spines are centred at 0 $x'$-coordinate. All simulations are analysed until the flux rope reaches a height of $y=600\,\mathrm{Mm}$ or $t=4000\,\mathrm{s}$, whichever comes first. 

\subsubsection{Class I}
\noindent
{\bf Eruptive cases}
\vspace{0.25cm}

\begin{figure}
    \centering
    \includegraphics[width=1\linewidth]{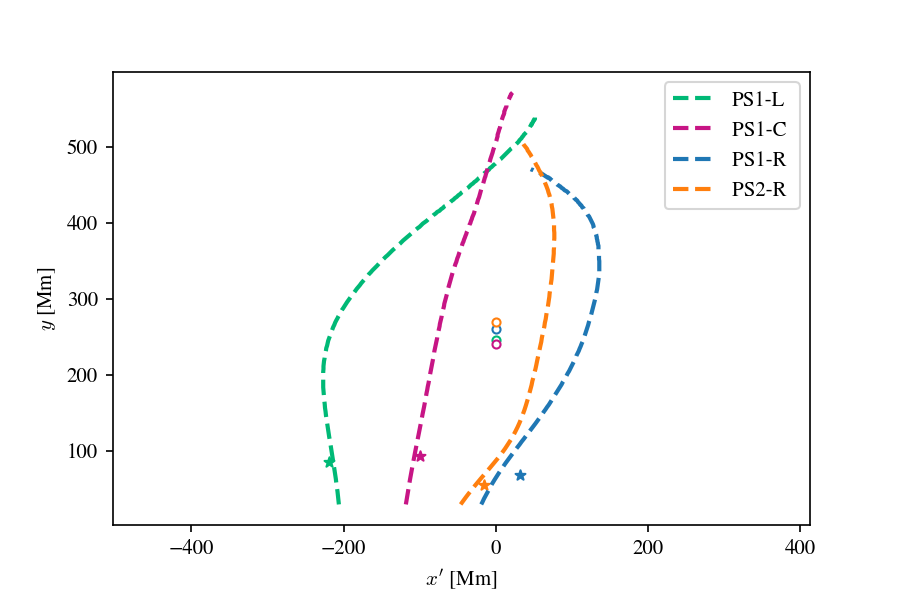}
    \caption{FR trajectories for the class I$_\mathrm{e}$ cases.
    Dashed lines represent the path followed by the FR, stars the LNP location, and circles the GNP position for each case. }
    \label{fig:trajectory-ein}
\end{figure}

Cases PS1-L, PS1-C, PS1-R and PS2-R belong to class I eruptive cases (see top row panels and middle-row right panel in Fig.~\ref{f:scenarios}). All their LNPs are inside the PS. For the cases PS1-L, PS1-C and PS1-R we consider the same magnetic configuration of the PS but different horizontal distances between the PS spine and the FR (see $x_\text{PS}$ in Table~\ref{t:PSparameters}). These relative distances determine different $(x,y)$ positions of the null points. 
Figure~\ref{fig:trajectory-ein} shows the trajectory of the FR for the different eruptive cases. The dashed-line of a given colour represents the FR trajectory, the stars of the same colour indicate the location of its LNP, and the circles indicate the position of its GNP. 

There is a common behaviour for class I eruptions. Initially, the FR moves towards the LNP, therefore the location of the LNP determines the direction of the initial deflection. After that, the LNP is deformed by the displacement of the FR, and the latter continues to rise towards the new direction of low magnetic energy. The FR is guided towards the PS spine, located above the GNP. The final directions of all class I$_\mathrm{e}$ trajectories eventually converge to a path parallel to the PS spine (which is not always radial). The arrival time and speed at this path will depend on the previous trajectory induced by the LNP. For example, FRs whose initial trajectory is more aligned with the direction of the GNP will eject faster (e.g. PS1-C) than those that are deviated by the LNP in an opposite direction to that of the GNP (e.g. PS1-R).

\vspace{0.25cm}
\noindent
{\bf Non-eruptive cases}
\vspace{0.25cm}

\begin{figure}
    \centering
    \includegraphics[width=1\linewidth]{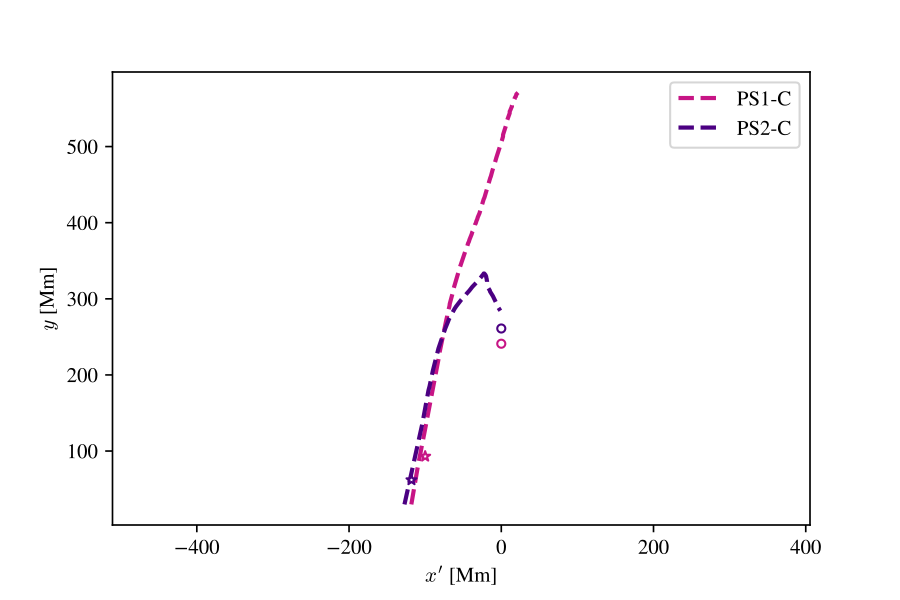}
    \caption{FR trajectories for the cases PS1-C (class I$_\mathrm{e}$) and PS2-C (class I$_\mathrm{ne}$). Dashed lines represent the path, stars the LNP location, and circles the GNP position for each case.  }
    \label{fig:trajectory-nein}
\end{figure}
Cases PS2-L and PS2-C are non-eruptive cases of class I (see left and centre panels in the middle row of Fig.~\ref{f:scenarios}).
PS2 cases have an almost identical morphology to PS1 cases (see top and middle rows of Fig.~\ref{f:scenarios}). The difference is that PS2 magnetic fields are twice stronger than PS1 cases. We compare here PS2-C (non-eruptive) with respect to PS1-C (eruptive); the description is analogous for PS2-L and PS1-L.

\begin{figure}
    \centering
     \includegraphics[width=1\linewidth]{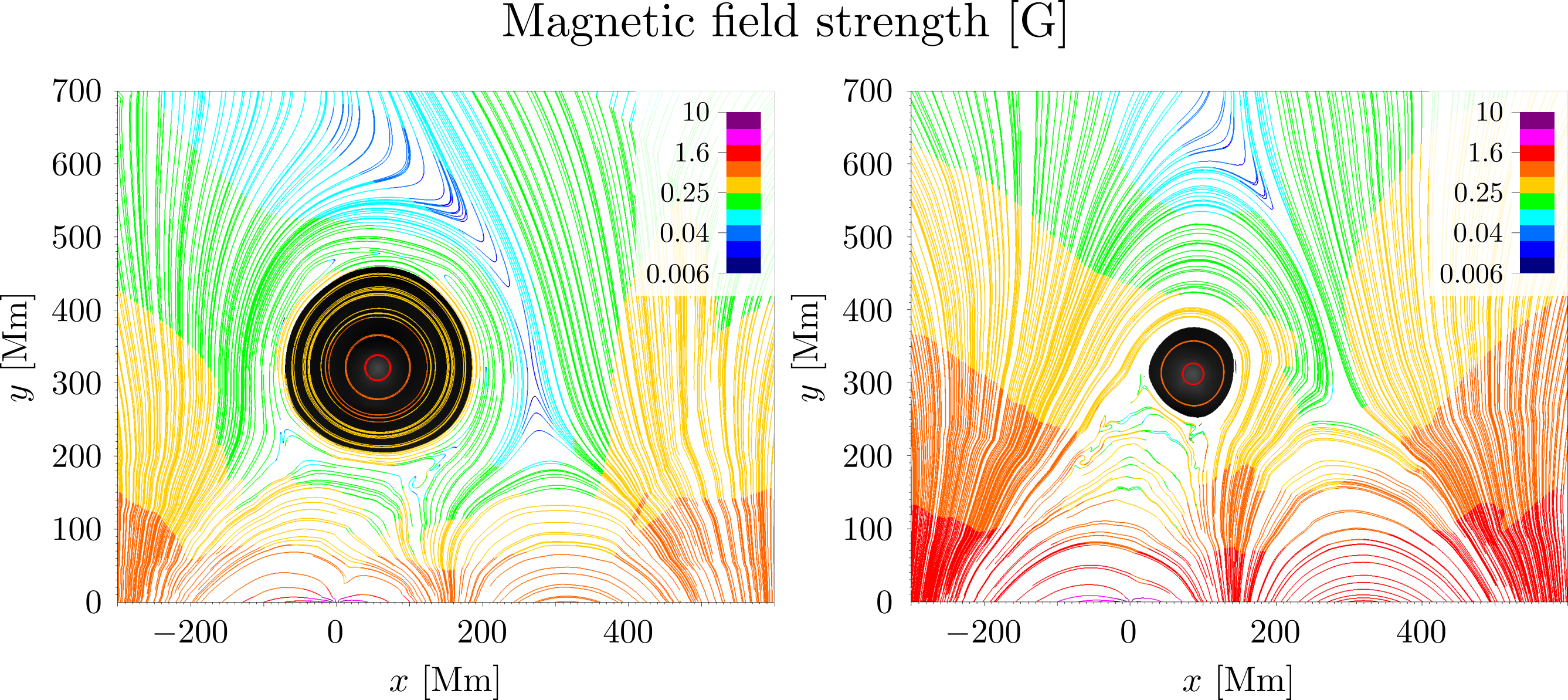}
    \caption{FR position (dark region) and magnetic field lines strength of PS1-C (eruptive; left panel) and PS2-C (non-eruptive; right panel) cases for $t=1400$ s. The animated evolution of the right panel is available in the HTML version.  }
    \label{fig:mag_cases}
\end{figure}

Figure~\ref{fig:trajectory-nein} shows the trajectory of PS1-C and PS2-C cases for comparison. The initial trajectories are remarkably similar, however at some point PS2-C case starts a decaying phase and consequently no eruption occurs. The FR continues its descendent motion towards the initial position of the GNP. To understand this behaviour we display in Figure~\ref{fig:mag_cases} both cases at time $t=1400$ s. The dark region highlights the FR location and the colour scale indicates the strength of the magnetic field lines. It can be seen that the volume of the FR for PS1-C case is larger than for PS2-C.
Studying the environment of the FR, we note that the magnetic cage, formed by the set of magnetic field lines from the PS lobe that overlay the FR and confine it, is wider and stronger for PS2-C. The differences between the two cages are most evident in the upper section of the cage, where the higher number of lines constituting the PS2-C cage is clear and their strengths can be compared by the colour levels. In addition, we observe a marked difference in the response of both cages to the FRs rise. In case PS1-C, we observe that the cage adapts and follows the shape of the FR, adopting a lock-like shape. In contrast, the PS2-C cage is not sufficiently prone to deformation, and this produces a noticeable imbalance between the fields supporting the FR below and those confining it above. Then, it is reasonable to infer that this produces a strong magnetic pressure gradient pushing the FR towards the base of the corona, which could trigger the decaying phase. The rigidity of the PS2-C cage may also be the reason for the reduced expansion of the FR, as can be noticed in the animation of the right panel of Fig.~\ref{fig:mag_cases} available in the HTML version. The animation shows the FR rising (until $t=2000\,$s) and its subsequent descent. At this last stage, the FR suffers a draining that results in the formation of detached magnetic islands around the FR boundaries. A similar behaviour is observed when comparing PS1-L and PS2-L cases, which share the same path until PS2-L slows its upward motion and finally starts the decaying phase. PS2-L also barely expands and suffers from mass draining.

Summarising, the general behaviour of Class I non-eruptive events is characterised by a rising and a decaying phase. During the rising phase, the FR trajectories present the same behaviour than the Class I eruptive events described above. The magnetic cages of non-eruptive cases withstand the upward motion and slow down the FR until the decaying phase begins. During the decaying phase, the FR no longer resists the action of gravity and is guided by the ambient magnetic field lines towards the chromosphere. 
In addition, non-eruptive FRs expand weakly under the pressure of the ambient magnetic field and they become smaller as their outer parts split into detached magnetic islands, sometimes completely destroying the identity of the FR.

\subsubsection{Class O}

\begin{figure}
    \centering
    \includegraphics[width=1\linewidth]{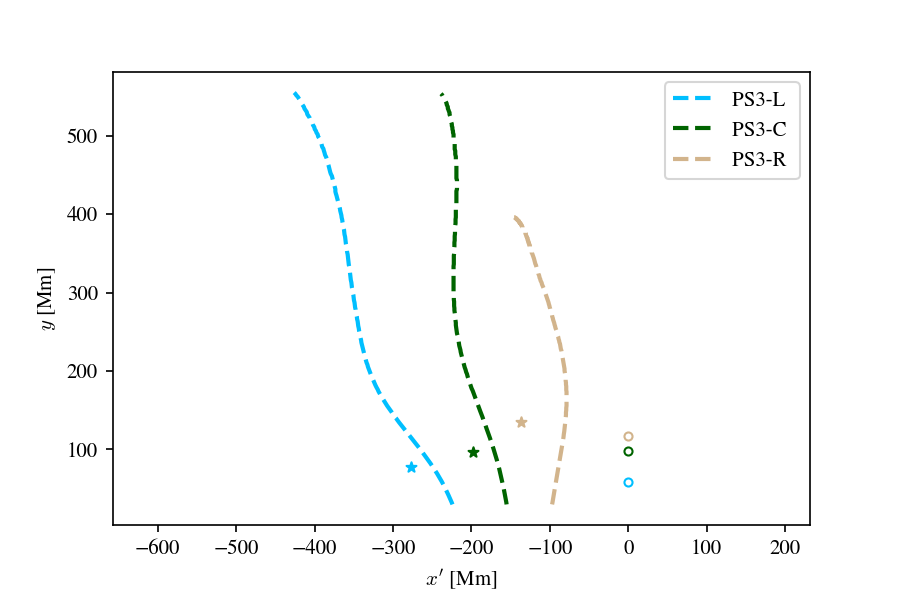}
    \caption{FR trajectories for the class O cases: PS3-R, PS3-C and PS3-L. Dashed lines represent the path, stars the LNP location, and circles the GNP position for each case.}
    \label{fig:trajectory-eout}
\end{figure}

The triad PS3-L, PS3-C and PS3-R belongs to the class O events. They have a common PS structure, but the FR horizontal position is different for each of them. Because these cases are topologically different from the class I cases (see bottom row of Fig.~\ref{f:scenarios}), the FR trajectories are not affected in the same way by the null points. This triad has their own spine-like structure whose base is located at the LNP. Figure~\ref{fig:trajectory-eout} shows the trajectories for the PS3 cases. It can be seen that PS3-L and PS3-C are barely affected by the GNP, heading initially towards their LNP and then continuing upwards into their own spine zone, guided by the open magnetic field lines of this spine. However, PS3-R case, whose initial position is almost equidistant from both null points, travels between them before reaching the spine of the LNP. 

Summarising, the FR trajectories for class O are initially headed towards their LNP and then continue upwards toward their own spine-like zone. The events are not influenced by the GNP except when the FR is relatively close to it. Moreover, simulations in this scenario always erupt, seemingly because the LNP is directly connected to the open field lines and there is no magnetic cage above the FR.

\subsection{Quantitative analysis}
\label{ss:quantitative}

\begin{figure}
    \centering
    \includegraphics[width=1\linewidth]{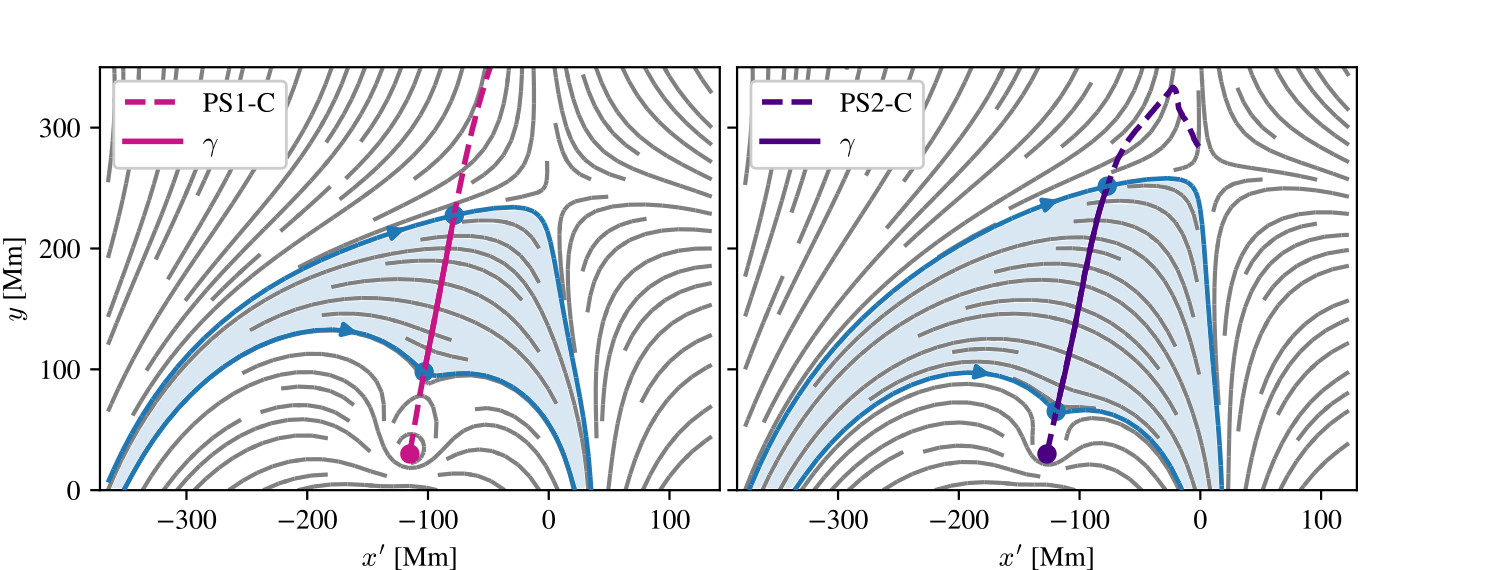}
    \caption{Magnetic field lines for events PS1-C (left) and PS2-C (right). The shaded light blue areas represent the magnetic cage above the FR, magenta and indigo dots indicate the initial position for PS1-C and PS2-C, respectively. The solid line represents the internal trajectory of the FR through which the flux of the magnetic cage is quantified.}
    \label{fig:mag_cagePS1-2}
\end{figure}
\begin{figure}
    \centering
    \includegraphics[width=1\linewidth]{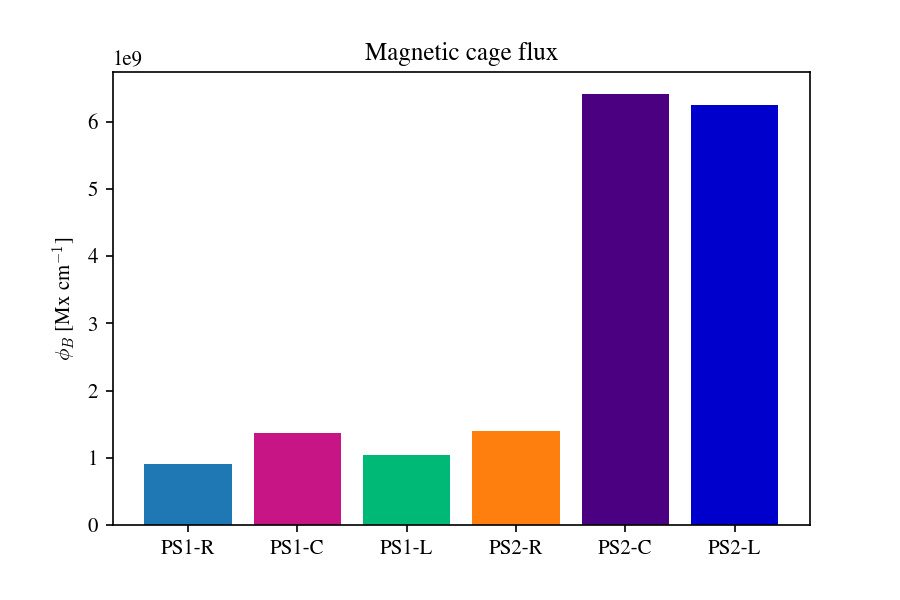}
    \caption{Total unsigned magnetic flux per unit length of the magnetic cage for PS1 and PS2 cases.}
    \label{fig:mag_flux}
\end{figure}

In this section we will focus on the magnetic cage and its effect on FR evolution. To be precise, we define a magnetic cage as the structure formed by all the field lines enclosing the FR and whose both ends are attached to the base of the corona (height $y=0$).
As suggested previously by Fig.~\ref{fig:mag_cases}, the magnetic cage of the non-eruptive events is larger and more intense than that of the eruptive cases. Figure~\ref{fig:mag_cagePS1-2} shows the magnetic cages for the PS1-C and PS2-C cases shaded in light blue. Inspired by the results of \citet{Li2020ApJ...900..128L,Li2021ApJ...917L..29L,Li2022ApJ...926L..14L}, we also determine the total unsigned magnetic flux to quantify the strength of the magnetic cages. Taking advantage of the symmetry considered in the $z$-direction, we calculate the initial magnetic flux per unit length $\phi_B$ through a path outlined by the FR trajectory as follows:
\begin{equation}
    \phi_B =\tfrac{1}{L_z} \int_A |\vec{B}\cdot\vec{dA}| = \int_\gamma |B_\perp|\, dS,
\end{equation}
where $B_\perp$ is the magnetic field transverse to a curve $\gamma$ defined by the FR path (denoted by the solid coloured lines in Fig.~\ref{fig:mag_cagePS1-2}). Figure~\ref{fig:mag_flux} shows the total unsigned magnetic flux of each magnetic cage for all PS1 and PS2 cases (PS3 cases do not produce magnetic cages). Note that the magnetic flux values for the non-eruptive cases (PS2-L and PS2-C) are remarkably large in comparison to the eruptive cases.

\begin{figure}
    \centering
    \includegraphics[width=0.9\linewidth]{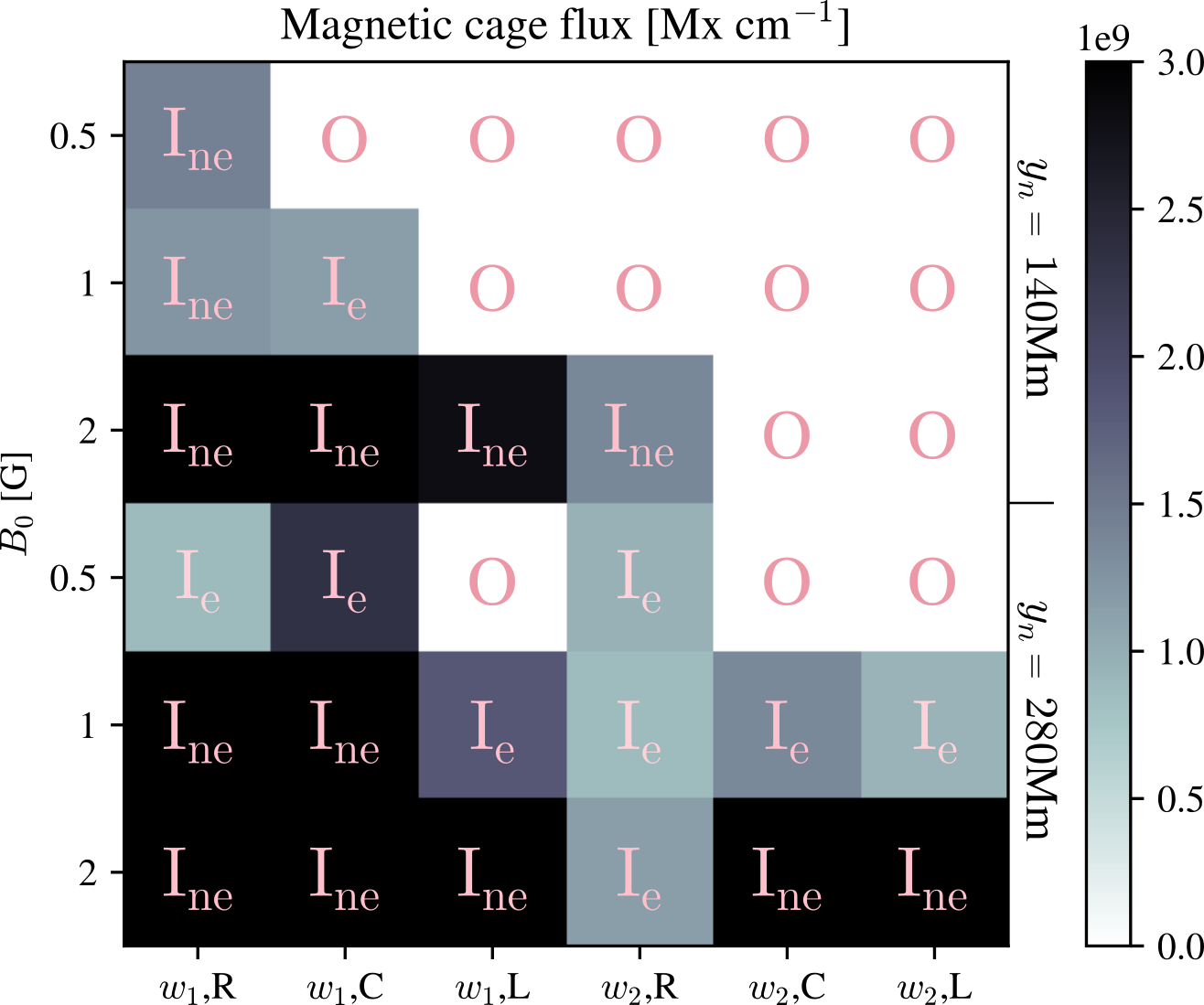}
    \caption{Parameter map for the simulation set. The colour bar represents the magnetic cage flux strength. $B_0$ parameter is the background magnetic field related to the PS, $w_1$ and $w_2$ refer to the PS lobe widths, R-C-L are the FR alignments, and $y_n$ is the PS null point height. For each box the dynamical behaviour is indicated.}
    \label{fig:fluxmap}
\end{figure}
To understand how the dynamical behaviour is affected by the simulation parameters, we include in Figure~\ref{fig:fluxmap} the magnetic cage flux (represented by the colours of the colour bar) for the whole simulation set, as a function of $B_0$ ($y$-axis) and width-alignment ($x$-axis). The widths ($w$) are $w_1\sim y_n$ and $w_2\sim1.5 y_n$, being $y_n$ the height of the null point, and the alignment R-C-L as described in Section 2.3. We also separate the cases according to $y_n$ (top and bottom parts of the plot) with the values denoted on the right. For each case we indicate its classification as we defined in previous section.
We notice again the correlation between larger magnetic cage fluxes and non-eruptive cases. In addition, these cases are more related with stronger $B_0$ magnetic fields and narrower pseudostreamers ($w_1$). This is expected as these parameters influence the magnetic flux of the pseudostreamer lobe. However, the magnetic cage flux will also depend on the position and parameters of the FR, i.e. on how many lobe lines actually belong to the cage. 

\begin{figure}
    \centering
    \includegraphics[width=1\linewidth]{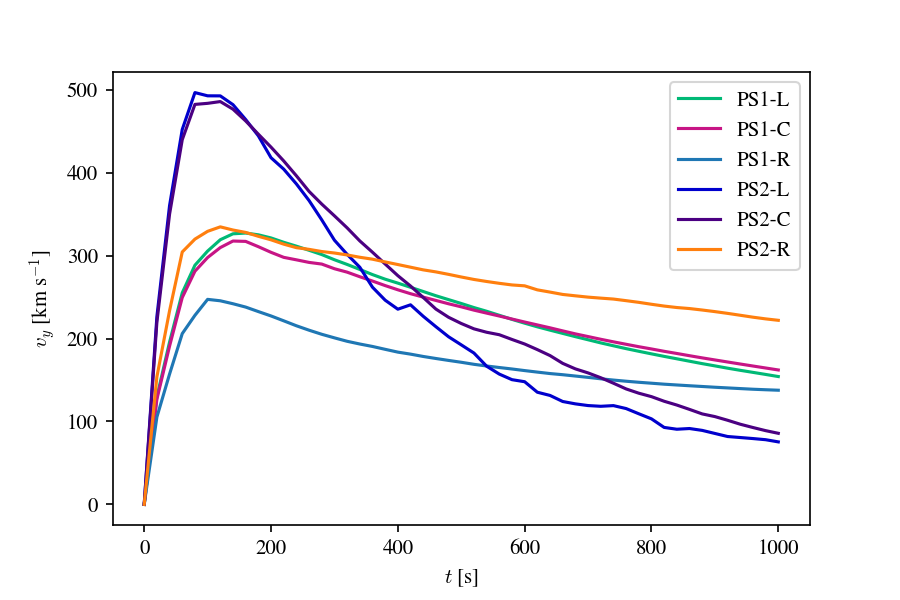}
    \caption{Vertical velocity $v_y$ as a function of time for PS1 and PS2 cases.}
    \label{fig:velocidad}
\end{figure}

We also analyse the evolution of the FR variables for PS1 and PS2 cases to understand how they are affected by the magnetic cage. PS3 cases are not included since they do not present a magnetic cage and, moreover, they follow the trends of the class I$_\mathrm{e}$.  
First, we analyse the evolution of FR velocity and total force in the $y$-direction. Figure~\ref{fig:velocidad} shows the vertical velocity curves ($v_y$) up to $t=1000~$s. The initial force for non-eruptive cases (PS2-L and PS2-C) is stronger and heads the FRs towards the LNP with notably higher speeds, likely due to the closer proximity to the LNP. However, after reaching the maximum value,  $v_y$ of I$_\mathrm{ne}$ events decreases more and steeper than $v_y$ of I$_\mathrm{e}$ events. Eventually $v_y$ becomes negative and the decaying phase of the FR starts. From the separate analysis of the force components (magnetic pressure and tension, gas pressure and gravity, not shown here), we find that the magnetic pressure gradient is the main responsible for the abrupt deceleration and the descent of the FR, after which gravity is the dominant decelerating force. This result is in agreement with the qualitative analysis presented in the previous section, in which we note that the concentration of field lines over the FR (see Fig.~\ref{fig:mag_cases}) seems to be responsible for exerting this magnetic pressure force. Gravity becomes the leading force once the FR is ``channelled'' by the lobe magnetic field lines (decaying phase).

\begin{figure}
    \centering
    \includegraphics[width=1\linewidth]{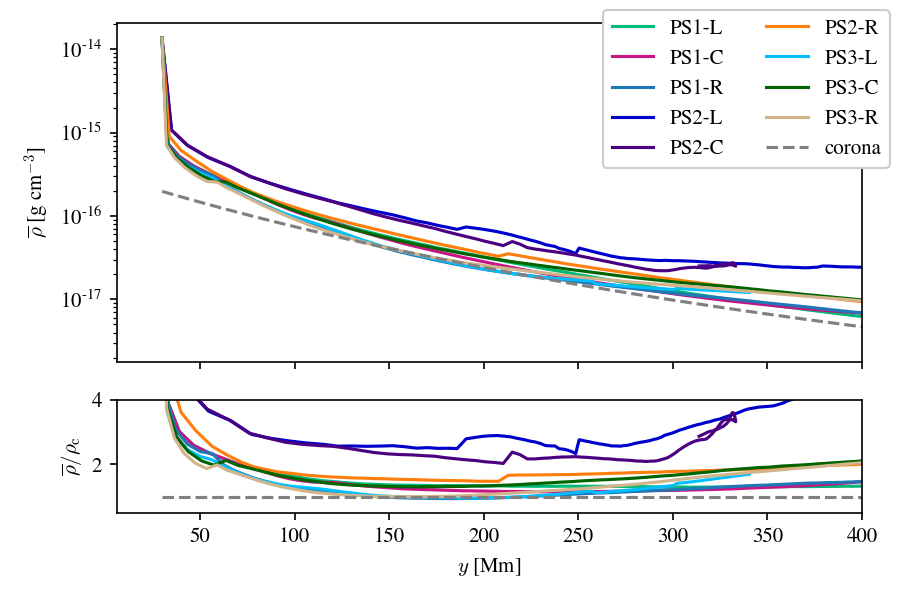}
    \caption{Average plasma density of the FR as a function of height. The dashed grey line represents the plasma density of the corona. The lower panel shows the corresponding ratio between the plasma density for  each case and the coronal plasma density $\rho_\mathrm{c}$. }
    \label{fig:dens}
\end{figure}

From the analysis of the FR variables, we note the major differences (between I$_\mathrm{e}$ and I$_\mathrm{ne}$ cases) in the evolution of the hydrodynamic ones. As we also mentioned in the previous section, the volume of the non-eruptive FRs remains small, contained by the strong magnetic cage surrounding them. Thus, the plasma density and gas pressure for non-eruptive FRs is higher than for the eruptive FRs, which manage to expand. Figure~\ref{fig:dens} shows the evolution of the FR average plasma density as a function of height, together with the coronal plasma density $\rho_\mathrm{c}$ (dashed grey line).  
Initially, all FRs are overdense and they quickly decrease their average density since they are out of external equilibrium. Afterwards, the density continues to decrease as the FRs expand, following the drop in ambient pressure with altitude. However, since the non-eruptive FRs (PS2-L and PS2-C) practically stop expanding, their densities tend asymptotically to a certain value. In the bottom panel of Fig.~\ref{fig:dens} we present the ratio between the FR and coronal density, which highlights the balance between the weight and the buoyant force. We note an important difference between eruptive and non-eruptive cases, while the former manage to reach densities similar to that of the corona, the non-eruptive ones remain at more than twice the coronal density due to the lack of expansion. Consequently, the buoyant force of these last cases is not strong enough to overcome the gravitational field and the action of the magnetic cage to produce the eruption.

\section{Discussion and Conclusions}

In this work we analyse the dynamic behaviour of a FR located near an isolated PS. The magnetic configuration produces the emergence of two magnetic null points associated with both structures: a LNP (local null point) formed by the cancellation of the FR and PS magnetic fields, and a GNP (global null point) related to the PS itself. We note that the LNP is determinant for the early evolution of the FR. All simulated cases show an initial deflection due to the attraction towards this point of low magnetic energy. The subsequent evolution depends on whether the FR is enclosed by the PS lobes (class I events) or not (class O), showing that the hierarchy of the null points depends on the topology. In class O, the LNP is associated to an intrinsic spine-like configuration and the FR is guided by its open magnetic field lines instead of travelling towards the PS spine.
For class I events, a second deflection can take place by the influence of the GNP, directing the FR to the PS spine. In this scenario, it is possible that the eruption fails. We determine that the magnetic cage, formed by the magnetic field lines from the PS lobe that encloses the FR, plays a crucial role in curbing the eruption. 

The non-eruptive cases, which initially reach higher velocities, are quickly decelerated by the magnetic cage. The cage field lines are compressed instead of adjusting to the rise of the FR, producing high magnetic pressure gradients that impulse the FR back to the surface. Also, we note for these cases that the expansion of the FR is inhibited by the magnetic cage, keeping it overdense and less buoyant, which helps to prevent the eruption.
Thus, we quantified the total unsigned magnetic fluxes of the cages, obtaining that in the non-eruptive cases the average value is almost six times higher than in the eruptive cases. This magnitude can be interpreted as a measure of the magnetic cage resistance. We also showed that cases with stronger magnetic field $B_0$ and narrower PS lobes are prone to be non-eruptive.

We show that the combination of a FR with a PS magnetic structure is topologically complex. Although the relative position between the FR and PS centre plays an essential role in predicting the non-radial motions of the FR trajectory, the magnetic flux contained in the magnetic cage seems to be the key parameter in determining whether an eruption can occur or not, in agreement with previous studies. Hence, we consider of utmost importance to attain improved magnetic field measurements such as those to be provided by missions like Solar Orbiter, PUNCH (Polarimeter to UNify the Corona and Heliosphere), and Aditya, among others, in order to analyse more observational events that can be compared with our results and to refine numerical models that contribute to space weather forecasts.

 \begin{acknowledgements}
      AS is doctoral fellow of CONICET. MC, GK, HC and AC are members of the Carrera del Investigador Cient\'ifico (CONICET). AS and MC acknowledge support from ANPCyT under grant number PICT No. 2016-2480. AS, MC, MVS and AC also acknowledge support by SECYT-UNC grant number PC No. 33620180101147CB. AS, MC, MVS and AC acknowledge support from PIP under grant number No. 11220200103150CO. MVS acknowledges support from the European Research Council (ERC) under the European Union's Horizon 2020 research and innovation programme (grant agreement No 724326). HC appreciates support from grants MSTCAME8181TC (UTN) and PIP 11220200102710CO (CONICET). Also, we thank the Centro de C\'omputo de Alto Desempe\~no (UNC), where the simulations were carried out. 
 \end{acknowledgements}

% WARNING
%-------------------------------------------------------------------
% Please note that we have included the references to the file aa.dem in
% order to compile it, but we ask you to:
%
% - use BibTeX with the regular commands:
%   \bibliographystyle{aa} % style aa.bst
%   \bibliography{Yourfile} % your references Yourfile.bib
%
% - join the .bib files when you upload your source files
%-------------------------------------------------------------------

\bibliographystyle{apalike}
\bibliography{ref}

\begin{thebibliography}{}

\bibitem[{Amari} et~al., 2018]{Amari2018}
{Amari}, T., {Canou}, A., {Aly}, J.-J., {Delyon}, F., and {Alauzet}, F. (2018).
\newblock {Magnetic cage and rope as the key for solar eruptions}.
\newblock {\em \nat}, 554(7691):211--215.

\bibitem[{Baumgartner} et~al., 2018]{Baumgartner2018}
{Baumgartner}, C., {Thalmann}, J.~K., and {Veronig}, A.~M. (2018).
\newblock {On the Factors Determining the Eruptive Character of Solar Flares}.
\newblock {\em \apj}, 853(2):105.

\bibitem[{Bi} et~al., 2013]{Bi2013}
{Bi}, Y., {Jiang}, Y., {Yang}, J., {Zheng}, R., {Hong}, J., {Li}, H., {Yang},
  D., and {Yang}, B. (2013).
\newblock {Analysis of the Simultaneous Rotation and Non-radial Propagation of
  an Eruptive Filament}.
\newblock {\em \apj}, 773(2):162.

\bibitem[{C{\'e}cere} et~al., 2020]{cecere2020}
{C{\'e}cere}, M., {Sieyra}, M.~V., {Cremades}, H., {Mierla}, M., {Sahade}, A.,
  {Stenborg}, G., {Costa}, A., {West}, M.~J., and {D'Huys}, E. (2020).
\newblock {Large non-radial propagation of a coronal mass ejection on 2011
  January 24}.
\newblock {\em Advances in Space Research}, 65(6):1654--1662.

\bibitem[{Chen} et~al., 2013]{Chen2013ApJ...778...70C}
{Chen}, H., {Ma}, S., and {Zhang}, J. (2013).
\newblock {Overlying Extreme-ultraviolet Arcades Preventing Eruption of a
  Filament Observed by AIA/SDO}.
\newblock {\em \apj}, 778(1):70.

\bibitem[{Cremades} et~al., 2006]{2006AdSpR..38..461C}
{Cremades}, H., {Bothmer}, V., and {Tripathi}, D. (2006).
\newblock {Properties of structured coronal mass ejections in solar cycle 23}.
\newblock {\em Advances in Space Research}, 38:461--465.

\bibitem[{Edmondson} et~al., 2010]{Edmondson10}
{Edmondson}, J.~K., {Antiochos}, S.~K., {DeVore}, C.~R., and {Zurbuchen}, T.~H.
  (2010).
\newblock {Formation and Reconnection of Three-dimensional Current Sheets in
  the Solar Corona}.
\newblock {\em \apj}, 718(1):72--85.

\bibitem[{Filippov}, 2021]{Filippov2021}
{Filippov}, B. (2021).
\newblock {Mass of prominences experiencing failed eruptions}.
\newblock {\em \pasa}, 38:e018.

\bibitem[{Filippov}, 2019]{Filippov2019}
{Filippov}, B.~P. (2019).
\newblock {Mass ejections from the solar atmosphere}.
\newblock {\em Physics Uspekhi}, 62(9):847--864.

\bibitem[{Forbes}, 1990]{1990JGR....9511919F}
{Forbes}, T.~G. (1990).
\newblock {Numerical simulation of a catastrophe model for coronal mass
  ejections}.
\newblock {\em Journal of Geophysical Research}, 95:11919--11931.

\bibitem[{Fryxell} et~al., 2000]{2000ApJS..131..273F}
{Fryxell}, B., {Olson}, K., {Ricker}, P., {Timmes}, F.~X., {Zingale}, M.,
  {Lamb}, D.~Q., {MacNeice}, P., {Rosner}, R., {Truran}, J.~W., and {Tufo}, H.
  (2000).
\newblock {FLASH: An Adaptive Mesh Hydrodynamics Code for Modeling
  Astrophysical Thermonuclear Flashes}.
\newblock {\em The Astrophysical Journal Supplement Series}, 131:273--334.

\bibitem[{Gopalswamy} et~al., 2009]{2009JGRA..114.0A22G}
{Gopalswamy}, N., {M{\"a}kel{\"a}}, P., {Xie}, H., {Akiyama}, S., and
  {Yashiro}, S. (2009).
\newblock {CME interactions with coronal holes and their interplanetary
  consequences}.
\newblock {\em Journal of Geophysical Research (Space Physics)}, 114:A00A22.

\bibitem[{Green} et~al., 2018]{Green2018}
{Green}, L.~M., {T{\"o}r{\"o}k}, T., {Vr{\v{s}}nak}, B., {Manchester}, W., and
  {Veronig}, A. (2018).
\newblock {The Origin, Early Evolution and Predictability of Solar Eruptions}.
\newblock {\em \ssr}, 214(1):46.

\bibitem[{Gronkiewicz} et~al., 2016]{Gronk2016IAUS..320.}
{Gronkiewicz}, D., {Mrozek}, T., {Ko{\l}oma{\'n}ski}, S., and
  {Chru{\'s}li{\'n}ska}, M. (2016).
\newblock {Searching for failed eruptions interacting with overlying magnetic
  field}.
\newblock In {Kosovichev}, A.~G., {Hawley}, S.~L., and {Heinzel}, P., editors,
  {\em Solar and Stellar Flares and their Effects on Planets}, volume 320,
  pages 221--223.

\bibitem[{Jiang} et~al., 2018]{Jiang2018}
{Jiang}, C., {Feng}, X., and {Hu}, Q. (2018).
\newblock {Formation and Eruption of an Active Region Sigmoid. II.
  Magnetohydrodynamic Simulation of a Multistage Eruption}.
\newblock {\em \apj}, 866(2):96.

\bibitem[{Jing} et~al., 2018]{Jing2018}
{Jing}, J., {Liu}, C., {Lee}, J., {Ji}, H., {Liu}, N., {Xu}, Y., and {Wang}, H.
  (2018).
\newblock {Statistical Analysis of Torus and Kink Instabilities in Solar
  Eruptions}.
\newblock {\em \apj}, 864(2):138.

\bibitem[{Karna} et~al., 2021]{Karna2021}
{Karna}, N., {Savcheva}, A., {Gibson}, S., {Tassev}, S., {Reeves}, K.~K.,
  {DeLuca}, E.~E., and {Dalmasse}, K. (2021).
\newblock {Magnetofrictional Modeling of an Erupting Pseudostreamer}.
\newblock {\em \apj}, 913(1):47.

\bibitem[{Kay} et~al., 2013]{Kay2013}
{Kay}, C., {Opher}, M., and {Evans}, R.~M. (2013).
\newblock {Forecasting a Coronal Mass Ejection's Altered Trajectory: ForeCAT}.
\newblock {\em \apj}, 775:5.

\bibitem[{Kay} et~al., 2015]{2015ApJ...805..168K}
{Kay}, C., {Opher}, M., and {Evans}, R.~M. (2015).
\newblock {Global Trends of CME Deflections Based on CME and Solar Parameters}.
\newblock {\em \apj}, 805:168.

\bibitem[{Li} et~al., 2021]{Li2021ApJ...917L..29L}
{Li}, T., {Chen}, A., {Hou}, Y., {Veronig}, A.~M., {Yang}, S., and {Zhang}, J.
  (2021).
\newblock {Magnetic Flux and Magnetic Nonpotentiality of Active Regions in
  Eruptive and Confined Solar Flares}.
\newblock {\em \apjl}, 917(2):L29.

\bibitem[{Li} et~al., 2020]{Li2020ApJ...900..128L}
{Li}, T., {Hou}, Y., {Yang}, S., {Zhang}, J., {Liu}, L., and {Veronig}, A.~M.
  (2020).
\newblock {Magnetic Flux of Active Regions Determining the Eruptive Character
  of Large Solar Flares}.
\newblock {\em \apj}, 900(2):128.

\bibitem[{Li} et~al., 2022]{Li2022ApJ...926L..14L}
{Li}, T., {Sun}, X., {Hou}, Y., {Chen}, A., {Yang}, S., and {Zhang}, J. (2022).
\newblock {A New Magnetic Parameter of Active Regions Distinguishing Large
  Eruptive and Confined Solar Flares}.
\newblock {\em \apjl}, 926(2):L14.

\bibitem[{Liewer} et~al., 2015]{2015SoPh..290.3343L}
{Liewer}, P., {Panasenco}, O., {Vourlidas}, A., and {Colaninno}, R. (2015).
\newblock {Observations and Analysis of the Non-Radial Propagation of Coronal
  Mass Ejections Near the Sun}.
\newblock {\em \solphys}, 290:3343--3364.

\bibitem[{Mei} et~al., 2012]{Mei2012}
{Mei}, Z., {Shen}, C., {Wu}, N., {Lin}, J., {Murphy}, N.~A., and {Roussev},
  I.~I. (2012).
\newblock {Numerical experiments on magnetic reconnection in solar flare and
  coronal mass ejection current sheets}.
\newblock {\em \mnras}, 425:2824--2839.

\bibitem[{M{\"o}stl} et~al., 2015]{Mostl2015}
{M{\"o}stl}, C., {Rollett}, T., {Frahm}, R.~A., {Liu}, Y.~D., {Long}, D.~M.,
  {Colaninno}, R.~C., {Reiss}, M.~A., {Temmer}, M., {Farrugia}, C.~J.,
  {Posner}, A., {Dumbovi{\'c}}, M., {Janvier}, M., {D{\'e}moulin}, P.,
  {Boakes}, P., {Devos}, A., {Kraaikamp}, E., {Mays}, M.~L., and {Vr{\v s}nak},
  B. (2015).
\newblock {Strong coronal channelling and interplanetary evolution of a solar
  storm up to Earth and Mars}.
\newblock {\em Nature Communications}, 6:7135.

\bibitem[{Panasenco} et~al., 2013]{2013SoPh..287..391P}
{Panasenco}, O., {Martin}, S.~F., {Velli}, M., and {Vourlidas}, A. (2013).
\newblock {Origins of Rolling, Twisting, and Non-radial Propagation of Eruptive
  Solar Events}.
\newblock {\em \solphys}, 287:391--413.

\bibitem[{Rachmeler} et~al., 2014]{2014ApJ...787L...3R}
{Rachmeler}, L.~A., {Platten}, S.~J., {Bethge}, C., {Seaton}, D.~B., and
  {Yeates}, A.~R. (2014).
\newblock {Observations of a Hybrid Double-streamer/Pseudostreamer in the Solar
  Corona}.
\newblock {\em \apjl}, 787(1):L3.

\bibitem[{Robertson} and {Priest}, 1987]{1987SoPh..114..311R}
{Robertson}, J.~A. and {Priest}, E.~R. (1987).
\newblock {Line-Tied Magnetic Reconnection}.
\newblock {\em \solphys}, 114(2):311--327.

\bibitem[{Sahade} et~al., 2021]{Sahade2021}
{Sahade}, A., {C{\'e}cere}, M., {Costa}, A., and {Cremades}, H. (2021).
\newblock {Polarity relevance in flux-rope trajectory deflections triggered by
  coronal holes}.
\newblock {\em \aap}, 652:A111.

\bibitem[{Sahade} et~al., 2020]{Sahade2020}
{Sahade}, A., {C{\'e}cere}, M., and {Krause}, G. (2020).
\newblock {Influence of Coronal Holes on CME Deflections: Numerical Study}.
\newblock {\em \apj}, 896(1):53.

\bibitem[{Shen} et~al., 2011]{shen2011}
{Shen}, C., {Wang}, Y., {Gui}, B., {Ye}, P., and {Wang}, S. (2011).
\newblock {Kinematic Evolution of a Slow CME in Corona Viewed by STEREO-B on 8
  October 2007}.
\newblock {\em \solphys}, 269:389--400.

\bibitem[{Sieyra} et~al., 2020]{Sieyra2020}
{Sieyra}, M.~V., {C{\'e}cere}, M., {Cremades}, H., {Iglesias}, F.~A., {Sahade},
  A., {Mierla}, M., {Stenborg}, G., {Costa}, A., {West}, M., and {D'Huys}, E.
  (2020).
\newblock {Analysis of Large Deflections of Prominence–CME Events during the
  Rising Phase of Solar Cycle 24}.
\newblock {\em \solphys}, 295:126.

\bibitem[{T{\"o}r{\"o}k} and {Kliem}, 2005]{Torok2005}
{T{\"o}r{\"o}k}, T. and {Kliem}, B. (2005).
\newblock {Confined and Ejective Eruptions of Kink-unstable Flux Ropes}.
\newblock {\em \apjl}, 630(1):L97--L100.

\bibitem[{T{\"o}r{\"o}k} et~al., 2011]{Torok2011}
{T{\"o}r{\"o}k}, T., {Panasenco}, O., {Titov}, V.~S., {Miki{\'c}}, Z.,
  {Reeves}, K.~K., {Velli}, M., {Linker}, J.~A., and {De Toma}, G. (2011).
\newblock {A Model for Magnetically Coupled Sympathetic Eruptions}.
\newblock {\em \apjl}, 739(2):L63.

\bibitem[{van Driel-Gesztelyi} and {Green}, 2015]{vanDriel2015}
{van Driel-Gesztelyi}, L. and {Green}, L.~M. (2015).
\newblock {Evolution of Active Regions}.
\newblock {\em Living Reviews in Solar Physics}, 12(1):1.

\bibitem[{Wang} et~al., 2020]{Wang2020JGRA}
{Wang}, J., {Hoeksema}, J.~T., and {Liu}, S. (2020).
\newblock {The Deflection of Coronal Mass Ejections by the Ambient Coronal
  Magnetic Field Configuration}.
\newblock {\em Journal of Geophysical Research (Space Physics)}, 125(8):e27530.

\bibitem[{Wang} et~al., 2015]{Wang2015}
{Wang}, R., {Liu}, Y.~D., {Dai}, X., {Yang}, Z., {Huang}, C., and {Hu}, H.
  (2015).
\newblock {The Role of Active Region Coronal Magnetic Field in Determining
  Coronal Mass Ejection Propagation Direction}.
\newblock {\em \apj}, 814(1):80.

\bibitem[{Wang} and {Zhang}, 2007]{WangyZhang2007}
{Wang}, Y. and {Zhang}, J. (2007).
\newblock {A Comparative Study between Eruptive X-Class Flares Associated with
  Coronal Mass Ejections and Confined X-Class Flares}.
\newblock {\em \apj}, 665(2):1428--1438.

\bibitem[{Wang}, 2015]{Wang2015PS}
{Wang}, Y.~M. (2015).
\newblock {Pseudostreamers as the Source of a Separate Class of Solar Coronal
  Mass Ejections}.
\newblock {\em \apjl}, 803(1):L12.

\bibitem[{Wyper} et~al., 2021]{wyper2021}
{Wyper}, P.~F., {Antiochos}, S.~K., {DeVore}, C.~R., {Lynch}, B.~J., {Karpen},
  J.~T., and {Kumar}, P. (2021).
\newblock {A Model for the Coupled Eruption of a Pseudostreamer and Helmet
  Streamer}.
\newblock {\em \apj}, 909(1):54.

\bibitem[{Yang} et~al., 2018]{Yang2018}
{Yang}, J., {Dai}, J., {Chen}, H., {Li}, H., and {Jiang}, Y. (2018).
\newblock {Filament Eruption with a Deflection of Nearly 90 Degrees}.
\newblock {\em \apj}, 862(1):86.

\bibitem[{Yang} et~al., 2015]{Yang2015ApJ}
{Yang}, J., {Jiang}, Y., {Xu}, Z., {Bi}, Y., and {Hong}, J. (2015).
\newblock {Interchange Reconnection Forced by the Filament Eruption Inside a
  Pseudo-streamer}.
\newblock {\em \apj}, 803(2):68.

\bibitem[{Zhang} et~al., 2001]{Zhang2001}
{Zhang}, J., {Dere}, K.~P., {Howard}, R.~A., {Kundu}, M.~R., and {White}, S.~M.
  (2001).
\newblock {On the Temporal Relationship between Coronal Mass Ejections and
  Flares}.
\newblock {\em \apj}, 559(1):452--462.

\bibitem[{Zuccarello} et~al., 2012]{Zuccarello2012}
{Zuccarello}, F.~P., {Bemporad}, A., {Jacobs}, C., {Mierla}, M., {Poedts}, S.,
  and {Zuccarello}, F. (2012).
\newblock {The Role of Streamers in the Deflection of Coronal Mass Ejections:
  Comparison between STEREO Three-dimensional Reconstructions and Numerical
  Simulations}.
\newblock {\em \apj}, 744:66.

\end{thebibliography}

\end{document}